\documentclass[12pt]{article}

\usepackage{graphicx}
\begin{document}

\begin{center}
{\bf Extended phase space thermodynamics of magnetized black holes with nonlinear electrodynamics} \\
\vspace{5mm} S. I. Kruglov
\footnote{E-mail: serguei.krouglov@utoronto.ca}
\underline{}
\vspace{3mm}

\textit{Department of Physics, University of Toronto, \\60 St. Georges St.,
Toronto, ON M5S 1A7, Canada\\
Canadian Quantum Research Center, \\
204-3002 32 Ave., Vernon, BC V1T 2L7, Canada} \\
\vspace{5mm}
\end{center}
\begin{abstract}
Einstein's gravity in AdS space coupled to nonlinear electrodynamics (NED) with two parameters is studied. We investigate magnetically charged black holes. The metric and mass functions and their asymptotic are obtained showing that black holes may have one or two horizons. Thermodynamics in extended phase space was studied and it was proven that the first law of black hole thermodynamics and the generalized Smarr relation hold. The magnetic potential and the vacuum polarization conjugated to coupling (NED parameter), are computed and depicted.
We calculate the Gibbs free energy and heat capacity.
\end{abstract}



It is known that black holes are thermodynamic systems \cite{Bardeen,Jacobson,Padmanabhan} with entropy and temperature \cite{Bekenstein,Hawking}. Black holes phase transitions were studied by Page and Hawking in Schwarzschild-AdS space-time \cite{Page}. The importance of gravity in Anti de Sitter (AdS) space-time is because of the holographic principle) \cite{Maldacena} which has applications in condensed matter physics. In an extended phase space black hole thermodynamics the negative cosmological constant plays the role of a thermodynamic pressure which is conjugated to volume \cite{Dolan,Kubiznak,Teo,Mann1}. This allows to proof the first law of black hole thermodynamics. Here, we study Einstein-AdS theory coupled to nonlinear electrodynamics (NED) to smooth out singularities which are present in the linear Maxwell electrodynamics. M. Born and L. Infeld \cite{Born} firstly proposed NED which does not have a singularity of point-like charges and have the electric field energy finite. For weak fields Born--Infeld (BI) electrodynamics is converted into Maxwell’s theory.
In this letter we explore NED \cite{Kruglov,Kruglov1} with two parameters that have similar features as BI electrodynamics, the absence of singularities. We will study magnetic black holes as electric black holes with weak-field Maxwell's limit possess singularities \cite{Bronnikov}.

The Einstein's gravity in AdS space-time is described by the action
\begin{equation}
I=\int d^{4}x\sqrt{-g}\left(\frac{R-2\Lambda}{16\pi G_N}+\mathcal{L}(\mathcal{F}) \right),
\label{1}
\end{equation}
where $\Lambda=-3/l^2$ is the negative cosmological constant and $l$ is the AdS radius. Here, we explore the source of gravity  NED with the Lagrangian \cite{Kruglov,Kruglov1}
\begin{equation}
{\cal L}=-\frac{{\cal F}}{4\pi\left(1+\epsilon(2\epsilon\beta {\cal F})^\gamma\right)}.
\label{2}
\end{equation}
${\cal F}=F^{\mu\nu}F_{\mu\nu}/4=(B^2-E^2)/2$, and $E$ and $B$ being the electric and magnetic fields, correspondingly, $\epsilon=\pm 1$,  $\beta>0$, $\gamma>0$. If $B>E$ one uses $\epsilon=1$ \cite{Kruglov} and when $B<E$ we put $\epsilon=-1$. For $\epsilon=-1$ the electric field at the origin and the electrostatic energy are finite \cite{Kruglov1}. At the weak-field limit Lagrangian (2) becomes the Maxwell's Lagrangian. At $\gamma=1$ in Eq. (2) we have the Lagrangian of rational NED \cite{Kruglov2}.
The Einstein's and field equations follow from action (1),
\begin{equation}
R_{\mu\nu}-\frac{1}{2}g_{\mu \nu}R+\Lambda g_{\mu \nu} =8\pi G_N T_{\mu \nu},
\label{3}
 \end{equation}
\begin{equation}
\partial _{\mu }\left( \sqrt{-g}\mathcal{L}_{\mathcal{F}}F^{\mu \nu}\right)=0,
\label{4}
\end{equation}
with $\mathcal{L}_{\mathcal{F}}=\partial \mathcal{L}( \mathcal{F})/\partial \mathcal{F}$. The energy-momentum tensor is given by
\begin{equation}
 T_{\mu\nu }=F_{\mu\rho }F_{\nu }^{~\rho }\mathcal{L}_{\mathcal{F}}+g_{\mu \nu }\mathcal{L}\left( \mathcal{F}\right).
\label{5}
\end{equation}
We will study spherical symmetrical solutions of Einstein's equation (3) with the line element squared
\begin{equation}
ds^{2}=-f(r)dt^{2}+\frac{1}{f(r)}dr^{2}+r^{2}\left( d\theta
^{2}+\sin ^{2}(\theta) d\phi ^{2}\right).
\label{6}
\end{equation}
Let us consider black holes as a magnetic monopole possessing the magnetic field $B=q/r^2$ and magnetic charge $q$.
The metric function can be found as \cite{Bronnikov}
\begin{equation}
f(r)=1-\frac{2m(r)G_N}{r},
\label{7}
\end{equation}
where the mass function is given by
\begin{equation}
m(r)=m_0+4\pi\int\rho (r)r^{2}dr.
\label{8}
\end{equation}
$m_0$ is the Schwarzschild mass (an integration constant), and $\rho$ is the energy density.
From Eq. (5) the magnetic energy density with the energy density due to AdS space-time is given by
\begin{equation}
\rho=\frac{q^2r^{4(\gamma-1)}}{8\pi \left[r^{4\gamma}+(q^2\beta)^\gamma\right]}-\frac{3}{8\pi G_Nl^2}.
\label{9}
\end{equation}
Making use of Eqs. (8) and (9) one obtains the mass function
\begin{equation}
m(r)=m_0+\frac{q^2r^{4\gamma-1}}{2\alpha(4\gamma-1)} F\left(1,1-\frac{1}{4\gamma};2-\frac{1}{4\gamma};-\frac{r^{4\gamma}}{\alpha}\right)-\frac{r^3}{2G_Nl^2},
\label{10}
\end{equation}
where $\alpha=(q^2\beta)^\gamma$ and $F(a,b;c;z)$ is the hypergeometric function. By virtue of Eqs. (7) and (10) we find the metric function
\begin{equation}
f(r)=1-\frac{2m_0 G_N}{r}-\frac{q^2r^{4\gamma-2}G_N}{\alpha(4\gamma-1)} F\left(1,1-\frac{1}{4\gamma};2-\frac{1}{4\gamma};-\frac{r^{4\gamma}}{\alpha}\right)+\frac{r^2}{l^2}.
\label{11}
\end{equation}
To obtain the asymptotic of the metric function we use the formula \cite{Abramowitz} as $z\rightarrow 0$
\begin{equation}
F(a,b;c;z)=1+\frac{ab}{c}z+\frac{a(a+1)b(b+1)}{c(c+1)}z^2+....
\label{12}
\end{equation}
Then we find the asymptotic as $r\rightarrow 0$, when the Schwarzschild mass is zero ($m_0=0$),
\begin{equation}
f(r)=1+\frac{r^2}{l^2}-\frac{G_Nq^2r^{4\gamma-2}}{\alpha(4\gamma-1)}+\frac{G_Nq^2r^{8\gamma-2}}{\alpha^2(8\gamma-1)}+{\cal O}(r^{12\gamma-2}),
\label{13}
\end{equation}
 and $\gamma\geq 1/2$.
From Eq. (13) we obtain $f(0)=1$ which is a necessary condition in order to have the space-time regular.
To find the asymptotic of the metric function as $r\rightarrow \infty$, one can use the transformation \cite{Abramowitz}
\[
F(a,b;c;z)=\frac{\Gamma(c)\Gamma(b-a)}{\Gamma(b)\Gamma(c-a)}(-z)^{-a}F\left(a,1-c+a;1-b+a;\frac{1}{z}\right)
\]
\begin{equation}
+\frac{\Gamma(c)\Gamma(a-b)}{\Gamma(a)\Gamma(c-b)}(-z)^{-b}F\left(b,1-c+b;1-a+b;\frac{1}{z}\right).
\label{14}
\end{equation}
Making use of Eqs. (11) and (14) we obtain
\[
f(r)=1-\frac{2M G_N}{r}+\frac{q^2G_N}{r^2} F\left(1,\frac{1}{4\gamma};1+\frac{1}{4\gamma};-\frac{\alpha}{r^{4\gamma}}\right)+\frac{r^2}{l^2},
\]
\begin{equation}
M=m_0+m_M,~~~~m_M=\frac{q^{3/2}\pi}{8\beta^{1/4}\gamma\sin(\pi/(4\gamma))}.
\label{15}
\end{equation}
Here, $m_M$ is the magnetic mass of a black hole and $M$ is the total black hole mass (the ADM mass). For particular cases of $\gamma$ ($\gamma=1/2,~3/4,~1,~1.5$) solutions of Eqs. (3) and (4) were found in \cite{Kruglov3,Kruglov4,Kruglov5,Kruglov6} which agree with the general solution (15) for arbitrary $\gamma$.
When $\Lambda=0$ ($l\rightarrow\infty$) and as $r\rightarrow \infty$ one finds from Eqs. (12) and (15)
\begin{equation}
f(r)=1-\frac{2MG_N}{r}+ \frac{q^2G_N}{r^2}+\mathcal{O}(r^{-3}).
\label{16}
\end{equation}
Equation (16) shows that black holes have corrections to the Reissner--Nordstr\"{o}m solution in the order of $\mathcal{O}(r^{-3})$.
In the limit $\beta\rightarrow 0$ the metric function (16) becomes the Reissner--Nordstr\"{o}m metric function. The plot of metric function (15) is depicted in Fig. 1 at $m_0=0$, $G_N=1$, $q=1$, $\beta=0.1$, $l=10$.
\begin{figure}[h]
\includegraphics {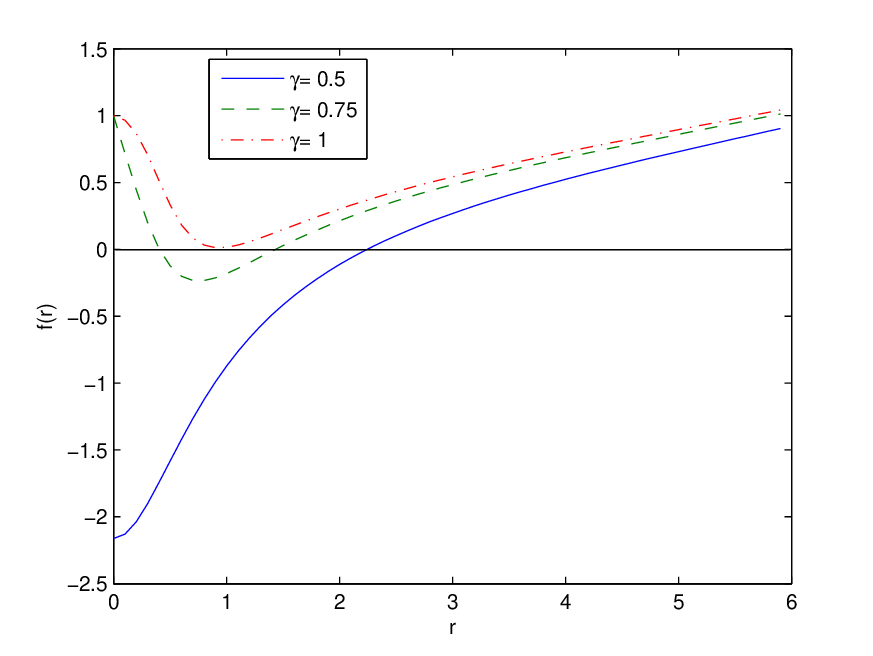}
\caption{\label{fig.1} The function $f(r)$ at $m_0=0$, $G_N=1$, $q=1$, $\beta=0.1$, $l=10$. Figure 1 shows that black holes could have one or two horizons.}
\end{figure}
According to Fig. (1), when parameter $\gamma$ increases the event horizon radius decreases. In accordance with Fig. 1 black holes may have one or two horizons.


In extended phase space thermodynamics the pressure is given by $P=-\Lambda/(8\pi)$  \cite{Kastor,Dolan1,Cvetic,Kubiznak1,Kubiznak2} and coupling $\beta$ is the thermodynamic value. The mass $M$ is treated as a chemical enthalpy, $M=U+PV$ and $U$ is the internal energy. With the help of the Euler's dimensional analysis with $G_N=1$  \cite{Smarr}, \cite{Kastor},
one finds dimensions $[M]=L$, $[S]=L^2$, $[P]=L^{-2}$, $[J]=L^2$, $[q]=L$, $[\beta]=L^2$ and
\begin{equation}
M=2S\frac{\partial M}{\partial S}-2P\frac{\partial M}{\partial P}+2J\frac{\partial M}{\partial J}+q\frac{\partial M}{\partial q}+2\beta\frac{\partial M}{\partial \beta},
\label{17}
\end{equation}
with $J$ being the black hole angular momentum.  The so-called vacuum polarization is the thermodynamic conjugate to coupling $\beta$  \cite{Teo} ${\cal B}=\partial M/\partial \beta $. The black hole entropy $S$, volume $V$ and pressure $P$ are given by
\begin{equation}
S=\pi r_+^2,~~~V=\frac{4}{3}\pi r_+^3,~~~P=-\frac{\Lambda}{8\pi}=\frac{3}{8\pi l^2}.
\label{18}
\end{equation} In the following we will study non-rotating black holes, $J=0$.
From Eq. (15) and equation $f(r_+)=0$, defining the event horizon radius $r_+$, we obtain
\begin{equation}
M(r_+)=\frac{r_+}{2G_N}+\frac{r_+^3}{2G_Nl^2}+\frac{q^2}{2r_+} F\left(1,\frac{1}{4\gamma};1+\frac{1}{4\gamma};-\frac{\alpha}{r_+^{4\gamma}}\right).
\label{19}
\end{equation}
By virtue of Eq. (19) at $G_N=1$, one finds
\[
dM(r_+)=\biggl[\frac{1}{2}+\frac{3r_+^2}{2l^2}- \frac{q^2}{2r_+^2}F\left(1,\frac{1}{4\gamma};1+\frac{1}{4\gamma};-\frac{\alpha}{r_+^{4\gamma}}\right)
\]
\[
+
 \frac{2\gamma \beta^\gamma q^{2\gamma+2}}{(4\gamma+1)r_+^{4\gamma+1}} F\left(2,1+\frac{1}{4\gamma};2+\frac{1}{4\gamma};-\frac{\alpha}{r_+^{4\gamma}}\right)\biggr]dr_+
-\frac{r_+^3}{l^3}dl
\]
\[
+\biggl[\frac{q}{r_+}F\left(1,\frac{1}{4\gamma};1+\frac{1}{4\gamma};-\frac{\alpha}{r_+^{4\gamma}}\right)
\]
\[
-\frac{\gamma\beta^\gamma q^{2\gamma+1}}{(4\gamma+1)r_+^{4\gamma+1}}F\left(2,1+\frac{1}{4\gamma};2+\frac{1}{4\gamma};-\frac{\alpha}{r_+^{4\gamma}}\right)\biggr]dq
\]
\begin{equation}
-\biggl[\frac{\gamma\beta^{\gamma-1} q^{2\gamma+2}}{2(4\gamma+1)r_+^{4\gamma+1}}F\left(2,1+\frac{1}{4\gamma};2+\frac{1}{4\gamma};-\frac{\alpha}{r_+^{4\gamma}}\right)\biggr]
d\beta,
\label{20}
\end{equation}
where we have used the relation \cite{Abramowitz}
\begin{equation}
\frac{dF(a,b;c;z)}{dz}=\frac{ab}{c}F(a+1,b+1;c+1;z).
\label{21}
\end{equation}
The Hawking temperature is defined by the relation
\begin{equation}
T=\frac{f'(r)|_{r=r_+}}{4\pi},
\label{22}
\end{equation}
where $f'(r)=\partial f(r)/\partial r$. Making use of Eqs. (15) and (22), one obtains the Hawking temperature
\[
T=\frac{1}{4\pi}\biggl[ \frac{1}{r_+}+\frac{3r_+}{l^2}-\frac{q^2}{r_+^3}F\left(1,\frac{1}{4\gamma};1+\frac{1}{4\gamma};-\frac{\alpha}{r_+^{4\gamma}}\right)
\]
\begin{equation}
+\frac{4\gamma\beta^{\gamma} q^{2\gamma+2}}{(4\gamma+1)r_+^{4\gamma+3}}F\left(2,1+\frac{1}{4\gamma};2+\frac{1}{4\gamma};-\frac{\alpha}{r_+^{4\gamma}}\right)\biggr].
\label{23}
\end{equation}
At the limit $\beta\rightarrow 0$ Eq. (23) becomes the Hawking temperature of Maxwell-AdS black hole.
With the help of Eqs. (18), (19) and (23) we obtain the first law of black hole thermodynamics
\begin{equation}
dM = TdS + VdP + \Phi dq + {\cal B}d\beta.
\label{24}
\end{equation}
Making use of Eqs. (20) with (24) one finds the magnetic potential $\Phi$ and the vacuum polarization ${\cal B}$ as follows:
\[
\Phi =\frac{q}{r_+}F\left(1,\frac{1}{4\gamma};1+\frac{1}{4\gamma};-\frac{\alpha}{r_+^{4\gamma}}\right)
\]
\[
-\frac{\gamma\beta^\gamma q^{2\gamma+1}}{(4\gamma+1)r_+^{4\gamma+1}}F\left(2,1+\frac{1}{4\gamma};2+\frac{1}{4\gamma};-\frac{\alpha}{r_+^{4\gamma}}\right),
\]
\begin{equation}
{\cal B}=-\frac{\gamma\beta^{\gamma-1} q^{2\gamma+2}}{2(4\gamma+1)r_+^{4\gamma+1}}F\left(2,1+\frac{1}{4\gamma};2+\frac{1}{4\gamma};-\frac{\alpha}{r_+^{4\gamma}}\right).
\label{25}
\end{equation}
The plots of $\Phi$ and ${\cal B}$ versus $r_+$ are given by Fig. 2.
\begin{figure}[h]
\includegraphics {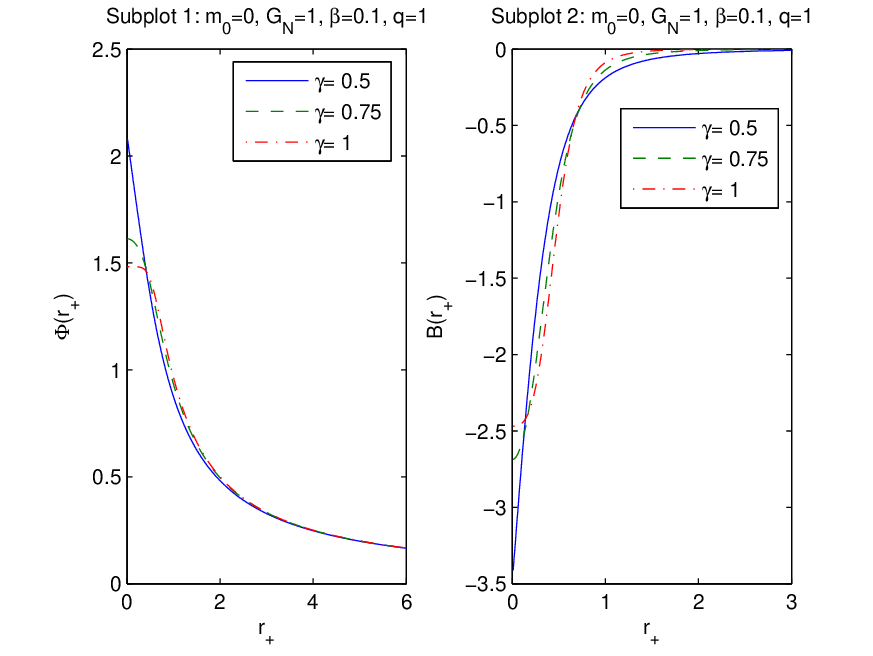}
\caption{\label{fig.2} The functions $\Phi$ and ${\cal B}$ vs. $r_+$ at $q=1$. The solid curve in subplot 1 is for $\gamma=0.5$, the dashed curve is for $\gamma=0.75$, and the dashed-doted curve is for $\gamma=1$. It follows that the magnetic potential $\Phi$ is finite at $r_+=0$ and becomes zero as $r_+\rightarrow \infty$. The function ${\cal B}$, in subplot 2, vanishes as $r_+\rightarrow \infty$ and is finite at $r_+=0$. }
\end{figure}
Figure 2, in left panel, shows that as $r_+\rightarrow \infty$ the magnetic potential vanishes ($\Phi(\infty)=0$), and at $r_+ = 0$ $\Phi$ is finite. When the parameter $\gamma$ increases, $\Phi(0)$ decreases.
According to Fig. 2 (right panel) at $r_+ = 0$ the vacuum polarization is finite and as $r_+\rightarrow \infty$, ${\cal B}$ becomes zero (${\cal B}(\infty)=0$). If the parameter $\gamma$ increases, ${\cal B}(0)$ also increases.
By virtue of Eqs. (18), (23) and (25) we obtain the generalized Smarr relation
\begin{equation}
M=2ST-2PV+q\Phi+2\beta{\cal B}.
\label{26}
\end{equation}


The local stability of black holes can be investigated by analyzing the heat capacity that is given by
\begin{equation}
C_q=T\left(\frac{\partial S}{\partial T}\right)_q=\frac{T\partial S/\partial r_+}{\partial T/\partial r_+}=\frac{2\pi r_+ T}{G_N\partial T/\partial r_+}.
\label{27}
\end{equation}
Making use of Eq. (23) we depicted the Hawking temperature in Fig. 3.
\begin{figure}[h]
\includegraphics {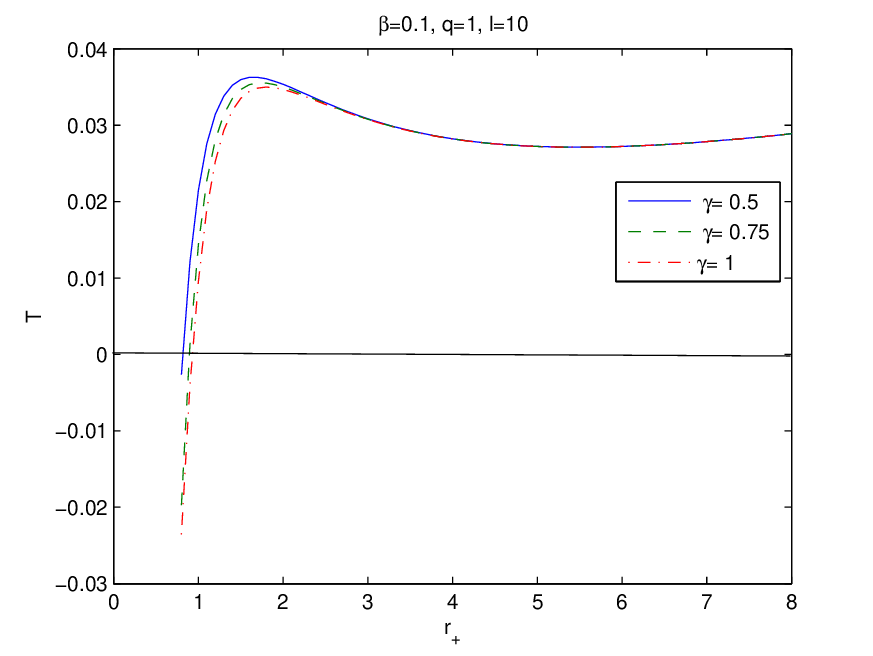}
\caption{\label{fig.3} The functions $T$ vs. $r_+$ at $q=1$, $\beta=0.1$, $l=10$. The solid curve in left panel is for $\gamma=0.5$, the dashed curve is for $\gamma=0.75$, and the dashed-doted curve is for $\gamma=1$. In some range of $r_+$ the Hawking temperature is negative where black holes do not exist. In the extremum of the Hawking temperature phase transitions take place.}
\end{figure}
In accordance with Eq. (27) when the Hawking temperature has an extremum the heat capacity possesses a singularity and the black hole phase transition occurs. For $\gamma=1$  black holes local stability was  analysed in \cite{Kruglov7}.
From Eq. (23) we obtain
\[
\frac{\partial T}{\partial r_+}=\frac{1}{4\pi}\biggl[-\frac{1}{r_+^2}+\frac{3}{l^2}+ \frac{3q^2}{r_+^4} F\left(1,\frac{1}{4\gamma};1+\frac{1}{4\gamma};-\frac{\alpha}{r_+^{4\gamma}}\right)
\]
\[
-\frac{16\gamma(\gamma+1) q^{2(\gamma+1)}\beta^\gamma}{(4\gamma+1)r_+^{4(\gamma+1)}} F\left(2,1+\frac{1}{4\gamma};2+\frac{1}{4\gamma};-\frac{\alpha}{r_+^{4\gamma}}\right)
\]
\begin{equation}
+\frac{32q^{4\gamma+2}\beta^{2\gamma}\gamma^2}{(8\gamma+1)r_+^{8\gamma+4}} F\left(3,2+\frac{1}{4\gamma};3+\frac{1}{4\gamma};-\frac{\alpha}{r_+^{4\gamma}}\right)\biggr].
\label{28}
\end{equation}
The heat capacity (27) is defined by Eqs. (23) and (28). The analyses of black holes local stability at parameters, $\gamma=0.5,~3/4,~1,~1.5$ were given in \cite{Kruglov3,Kruglov4,Kruglov5,Kruglov6}. One can do this for arbitrary $\gamma$ by using Eqs. (23), (27) and (28).

Making use of Eq. (23) we obtain the black hole equation of state (EoS)
\[
P=\frac{T}{2r_+}-\frac{1}{8\pi r_+^2}+\frac{q^2}{8\pi r_+^4}F\left(1,\frac{1}{4\gamma};1+\frac{1}{4\gamma};-\frac{\alpha}{r_+^{4\gamma}}\right)
\]
\begin{equation}
-\frac{\gamma\beta^{\gamma} q^{2\gamma+2}}{2\pi(4\gamma+1)r_+^{4\gamma+4}}F\left(2,1+\frac{1}{4\gamma};2+\frac{1}{4\gamma};-\frac{\alpha}{r_+^{4\gamma}}\right)\biggr].
\label{29}
\end{equation}
As $\beta\rightarrow 0$, Eq. (27) becomes EoS of charged Maxwell-AdS black hole \cite{Kubiznak1}. If the specific volume is defined as $v=2l_Pr_+$ ($l_P=\sqrt{G_N}=1$) \cite{Kubiznak1}, Eq. (27) is similar to the Van der Waals EoS. Replacing $v=2r_+$ into Eq. (27) one finds
\[
P=\frac{T}{v}-\frac{1}{2\pi v^2}+\frac{2q^2}{\pi v^4}F\left(1,\frac{1}{4\gamma};1+\frac{1}{4\gamma};-\frac{16^\gamma\alpha}{v^{4\gamma}}\right)
\]
\begin{equation}
-\frac{\gamma\beta^{\gamma}q^{2\gamma+2}2^{4\gamma+3}}{\pi(4\gamma+1)v^{4(\gamma+1)}}
F\left(2,1+\frac{1}{4\gamma};2+\frac{1}{4\gamma};-\frac{16^\gamma\alpha}{v^{4\gamma}}\right)\biggr].
\label{30}
\end{equation}
The plot of $P$ versus $v$ is depicted in Fiq. 4 for $q=\beta=1$, $T=0.05$.
\begin{figure}[h]
\includegraphics {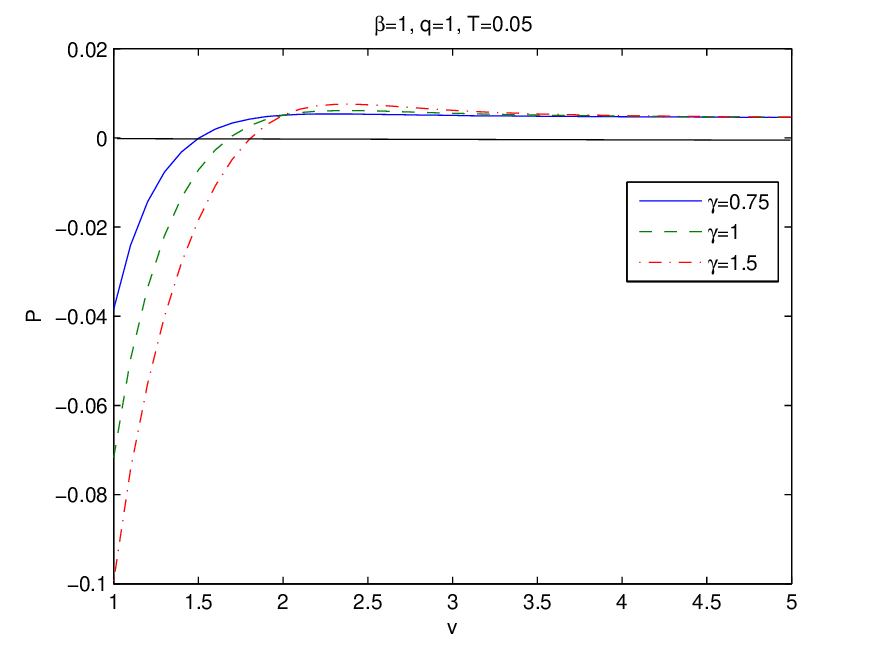}
\caption{\label{fig.4} The functions $P$ vs. $v$ at $q=\beta=1$, $T=0.05$. The solid curve for $\gamma=0.75$, the dashed curve is for $\gamma=1$, and the dashed-doted curve is for $\gamma=1.5$. }
\end{figure}
For some values of specific volume $v$ the pressure becomes negative (non-physical).
The critical points (inflection points) can be found by equations $\partial P/\partial v=0$, $\partial^2 P/\partial v^2=0$. It is impossible to find analytical solutions for critical points. Equations for critical points look cumbersome, so we will not present them here.
At the critical values $P-v$ diagrams for some parameters look similar to Van der Waals liquid diagrams possessing inflection points.

When $M$ is treated as a chemical enthalpy the Gibbs free energy is given by
\begin{equation}
G=M-TS.
\label{29}
\end{equation}
With the help of Eqs. (18), (19), and (23) we find
\[
G=\frac{r_+}{4}-\frac{2\pi r_+^3P}{3}+\frac{3q^2}{4r_+}F\left(1,\frac{1}{4\gamma};1+\frac{1}{4\gamma};-\frac{\alpha}{r_+^{4\gamma}}\right)
\]
\begin{equation}
-\frac{\gamma\beta^{\gamma}q^{2\gamma+2}}{(4\gamma+1)r_+^{4\gamma+1}}
F\left(2,1+\frac{1}{4\gamma};2+\frac{1}{4\gamma};-\frac{\alpha}{r_+^{4\gamma}}\right)\biggr].
\label{30}
\end{equation}
The plot of $G$ versus $T$ is depicted in Fig. 5 for $\beta=q=1$, $\gamma=3/8$.
\begin{figure}[h]
\includegraphics {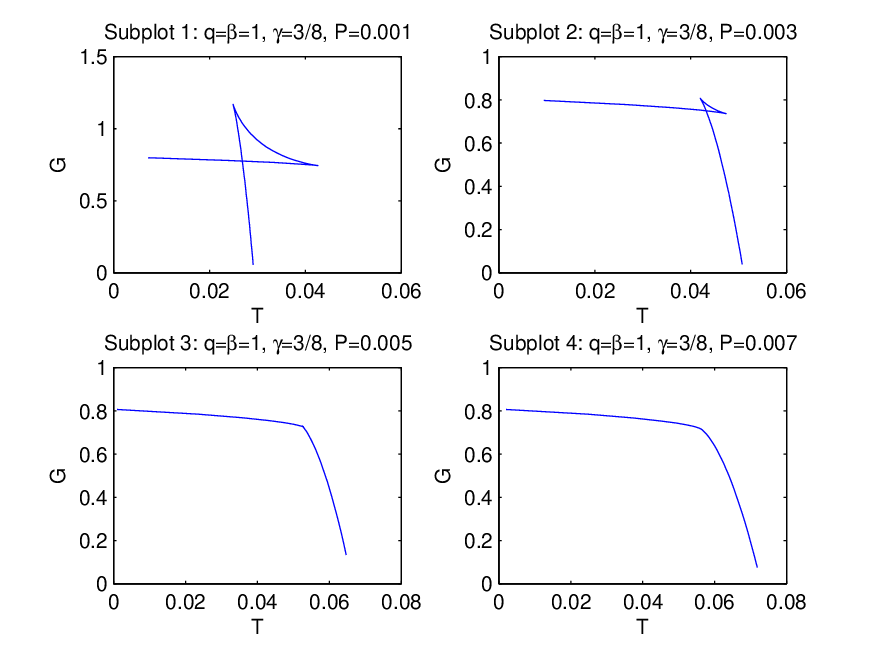}
\caption{\label{fig.5} The functions $G$ vs. $T$ at $q==\beta=1$, $\gamma=3/8$ for $P=0.001$, $P=0.003$, $P=0.005$ and $P=0.007$.
Subplots 1 and 2 show the critical 'swallowtail' behavior with first-order phase transitions between small and large black holes. Subplots 3  corresponds to the case of critical point where second-order phase transition occurs ($P_c\approx 0.005$). Subplots 4 shows  non-critical behavior of the Gibbs free energy.}
\end{figure}
For some parameters $\gamma=0.5,~3/4,~1,~1.5$, critical points and phase transitions of black holes were investigated in \cite{Kruglov3,Kruglov4,Kruglov5,Kruglov6} by analyzing the Gibbs free energy. One can study black holes phase transitions in our model for arbitrary $\gamma$ with the help of Gibbs's free energy (30).


The summary of results presented are as follows. Magnetic black hole solutions in Einstein-AdS gravity coupled to NED with two parameters are obtained. We obtain the metric and mass functions and their asymptotic with corrections to the Reissner--Nordstr\"{o}m solution. The the magnetic mass of a black hole is found. By plotted the metric function we found that black holes can have one or two horizons and when parameter $\gamma$ increases the event horizon radius decreases. We have studied the black holes thermodynamics in an extended phase space in AdS space-time. The thermodynamic quantity conjugated to coupling $\beta$ (the vacuum polarization), and thermodynamic potential, conjugated to magnetic charge, were obtained and plotted. We proofed that the first law of black hole thermodynamics and the generalized Smarr relation hold for any parameter $\gamma$. To analyse phase transitions we computed the Hawking temperature and Gibbs free energy. This allows us to make analyses of zero-order, first-order, and second-order phase transitions for arbitrary $\gamma$. Previous analyse of the model with particular case of $\gamma$ shows that black hole thermodynamics of our model is similar to the Van der Waals liquid–gas thermodynamics.

\end{document}